\documentclass[twocolumn]{autart}    

\usepackage{graphicx}          

\usepackage{natbib}
\usepackage[caption=false]{subfig}
\usepackage{epsfig}
\usepackage{epstopdf}
\usepackage{amsmath}
\usepackage{amssymb}
\usepackage{color}
\DeclareMathOperator*{\argmax}{arg\,max}

\usepackage{comment}

\begin{document}

\begin{frontmatter}

\title{Detection-averse optimal and receding-horizon control \\ for Markov decision processes\thanksref{footnoteinfo}} 

\thanks[footnoteinfo]{This paper was not presented at any IFAC
meeting. Corresponding author N.~Li. Email {\it nanli@umich.edu}}

\author[UM]{Nan Li}\ead{nanli@umich.edu},    
\author[UM]{Ilya Kolmanovsky}\ead{ilya@umich.edu},               
\author[UM]{Anouck Girard}\ead{anouck@umich.edu}  

\address[UM]{Department of Aerospace Engineering, University of Michigan, Ann Arbor, Michigan, USA}  


\begin{abstract}                          
In this paper, we consider a Markov decision process (MDP), where the ego agent has a nominal objective to pursue while needs to hide its state from detection by an adversary. After formulating the problem, we first propose a value iteration (VI) approach to solve it. To overcome the ``curse of dimensionality'' and thus gain scalability to larger-sized problems, we then propose a receding-horizon optimization (RHO) approach to obtain approximate solutions. We use examples to illustrate and compare the VI and RHO approaches, and to show the potential of our problem formulation for practical applications.
\end{abstract}

\end{frontmatter}

\section{Introduction}\label{sec:1}

Privacy is one of the ingredients of cyber-physical security, and has been gaining increasing attention in the age of cyber-physical systems and internet of things \citep{giraldo2017security}. For a dynamic system, privacy often corresponds to the observability/detectability of the system's state. In this paper, we consider privacy in the setting of controlled stochastic dynamic systems represented as Markov decision processes (MDPs) \citep{howard1960dynamic}.

In particular, we consider the decision-making process by an ego agent, who has a nominal objective to pursue while needs to hide its state from detection by an adversary. The detection by the adversary is through a stochastic observation function. This observation function provides partial/imperfect information about the ego agent's current state, used by the adversary to infer the value of the state. Specifically, the adversary's inference is, in general, a probability distribution on the state space, characterizing the adversary's degrees of believing that the current state is at each particular value. In this setting, to avoid detection by the adversary is to cause the adversary's belief in the true state value to be low.

Such a problem setting corresponds to many real-world situations. For example, an autonomous agent needs to carry out a task in an adversarial environment. The agent may want to hide its location from the adversary's sensing to protect itself from getting attacked.

Privacy for cyber-physical systems has been extensively investigated in the context of discrete-event systems (DES) \citep{lafortune2019discrete}. The notion in the language of DES that is most similar to the one discussed in this paper of avoiding detection by the adversary is called ``current-state opacity" \citep{jacob2016overview}. A DES is said to be current-state opaque if the adversary cannot determine {\it for sure} that the system is currently at a specific state/a specific set of states based on its limited observation. Although the notion and verification approaches of ``current-state opacity" have been recently extended to stochastic DES and to incorporate the concept of possibilities of being detected \citep{saboori2013current}, there are fundamental differences with the problem setting and solution approaches discussed in this paper: 1) The dynamics of a DES are typically event-triggered, while the dynamics of an MDP are time-based. 2) The analysis and control synthesis techniques in the context of DES are typically based on constructions of automatons, while those in the context of MDP, including the ones discussed in this paper, are essentially based on optimal control theory, e.g., dynamic programming.

In the context of MDP, a related problem is entropy maximization \citep{BIONDI2014384,savas2018entropy}. The entropy of a distribution quantifies the uncertainty of the event, and a larger entropy represents a higher degree of unpredictability. Differently from entropy maximization, the problem considered in this paper is only concerned with the value of the adversary's belief in the true state. For example, a belief distribution that assigns probability $1$ to a specific value has the minimum entropy $0$, representing a deterministic prediction and not being desired in entropy maximization. However, if this specific value with probability $1$ is not the true state value, such a belief distribution is desired in our problem setting, as it represents a prediction of the adversary that completely misses the ego agent's true state and thus is desired by the ego agent.

Another related problem is covariance control \citep{hotz1987covariance,chen2015optimal}. Similar to entropy, a ``larger'' covariance represents a higher degree of uncertainty. The same example as above can be used to illustrate the difference between covariance control and our problem setting.

To the best of our knowledge, the problem discussed in this paper has not been previously investigated in the literature. The contributions of this paper include, in addition to the formulation of the detection-averse MDP, the developments of a value iteration (VI) approach to solve it and a receding-horizon optimization (RHO) approach, which has better scalability than the one based on VI, to obtain approximate solutions. The proposed problem formulation also has a potential for real-world applications, which has been discussed above and will also be illustrated through an example in Section~\ref{sec:5}.

The remainder of this paper is organized as follows: In Section~\ref{sec:2}, we introduce the problem formulation. In Section~\ref{sec:3}, we describe an approach based on value iteration to solve the problem. To overcome the ``curse of dimensionality'' and gain scalability to  larger-sized problems, we propose an approach based on receding-horizon optimization to obtain approximate solutions in Section~\ref{sec:4}. We then use two examples to illustrate the problem and approaches in Section~\ref{sec:5}, and finally conclude the paper in Section~\ref{sec:6}.

The notations used in this paper are standard. In particular, we use $\mathbb{N}$ to denote the set of natural numbers (without $0$), and $\mathbb{N}_0 = \mathbb{N} \cup \{0\}$. We use $\mathbb{I}_{\Omega}(\omega)$ to denote the indicator function, $\mathbb{I}_{\Omega}(\omega) = 1$ if $\omega \in \Omega$ and $\mathbb{I}_{\Omega}(\omega) = 0$ otherwise, and $\mathbb{I}_{\omega'}(\omega) = \mathbb{I}_{\{\omega'\}}(\omega)$.

\section{Detection-averse Markov decision process}\label{sec:2}

\subsection{Markov decision process with nominal objective}

We consider a finite-space Markov decision process (MDP) for an ego agent represented by the following time-invariant state transition kernel:
\begin{equation}\label{equ:STK}
    p(x'|x,u) := \mathbb{P}(x_{t+1} = x'|x_t = x,u_t = u),\,\, \forall\, t \in \mathbb{N}_0,
\end{equation}
defined for all $x,x' \in X$ and $u \in U$, where $X$ and $U$ represent, respectively, a finite state space and a finite action space. In particular, it is assumed that the state $x_t$ is fully observable by the ego agent.

The nominal decision-making objective of the ego agent at time step $t$, for every nonnegative integer $t \in \mathbb{N}_0$, is represented using the expected value of an infinite-horizon discounted cumulative reward, as follows:
\begin{equation}\label{equ:Reward}
   \mathcal{R}(x_t) := \mathbb{E} \Big\{\sum_{\tau=0}^{\infty} \lambda^{\tau} R(x_{\tau|t},u_{\tau|t}) \big| x_{0|t} = x_t \Big\},
\end{equation}
where $R: X \times U \to \mathbb{R}$ is a stage reward function, and $\lambda \in (0,1)$ is a discount factor. By notation $(\cdot)_{\tau|t}$, we mean a predicted value of the variable $(\cdot)_{t+\tau}$ with the prediction made at time step $t$.

When solely pursuing this nominal objective, the ego agent maximizes \eqref{equ:Reward} subject
to the dynamics \eqref{equ:STK}, i.e.,
\begin{subequations}\label{equ:MDP}
\begin{align}
   \max_{\pi}\quad & \mathcal{R}^{\pi}(x_t), \\
   \text{s.t.}\quad &  p(x'|x,u),
\end{align}
\end{subequations}
where $\pi: X \to U$ is a policy that maps states to actions, and $\mathcal{R}^{\pi}(x_t)$ represents the value of \eqref{equ:Reward} under the policy $\pi$, i.e., $u_{\tau|t} = \pi(x_{\tau|t})$ for all $\tau \in \mathbb{N}_0$.

It is well-known that the solution to \eqref{equ:MDP} exists and satisfies the following Bellman equation (Theorems~6.2.6 and 6.2.10 of \cite{Puterman:1994:MDP:528623}):
\begin{align}\label{equ:Bellman}
    & V(x_t) \nonumber \\
    &= \max_{u \in U} \Big\{ R(x_t,u) + \lambda\, \mathbb{E} \big\{V(x_{1|t})| x_{0|t} = x_t, u_{0|t} = u \big\} \Big\}, \nonumber \\
    &= \max_{u \in U} \Big\{ R(x_t,u) + \lambda \sum_{x' \in X} p(x'|x_t,u) V(x') \Big\},
\end{align}
where $V(x):= \max_{\pi}\, \mathcal{R}^{\pi}(x)$ can be determined by the following value iteration:
\begin{equation}\label{equ:VI}
    V^{k+1}(x) = \max_{u \in U} \Big\{ R(x,u) + \lambda \sum_{x' \in X} p(x'|x,u) V^k(x') \Big\},
\end{equation}
which is convergent for all $x \in X$ as $k \to \infty$.

Once $V(x)$ is obtained, the optimal policy for the ego agent under nominal decision-making objective, $\pi^*$, can be obtained by
\begin{equation}\label{equ:Nominal_P}
    \pi^*(x) = \argmax_{u \in U} \Big\{ R(x,u) + \lambda \sum_{x' \in X} p(x'|x,u) V(x') \Big\}.
\end{equation}

In this paper, we assume that the optimal policy $\pi^*$ is unique. Such a uniqueness assumption for optimal policies of MDPs is often adopted in the literature (see \cite{cruz2004conditions} and references therein).

\subsection{Adversarial detection}

We consider a situation where an adversary is trying to detect the state of the ego agent, in particular, through the following observation kernel:
\begin{equation}\label{equ:OK}
    q(y|x) := \mathbb{P}(y_{t} = y|x_t = x),\quad \forall\, t \in \mathbb{N}_0,
\end{equation}
defined for all $x \in X$ and $y \in Y$, where $Y$ represents a finite observation space. It is assumed that the observation $y_{t}$ is conditionally independent of all other variables given $x_t$. In addition, we make the following assumptions about the adversary: 1) The adversary cannot observe the ego agent's action $u_t$ applied at every time step. However, 2) the adversary knows the state transition kernel \eqref{equ:STK} and the ego agent's nominal objective function \eqref{equ:Reward}.

The above assumptions are reasonable in many applications. On the one hand, the information asymmetry that the ego agent can fully observe its own state $x_t$ and action $u_t$ while the adversary can only obtain partial information $y_t$ exists in many situations, e.g., in cases where $x_t$ includes some internal states of the ego agent that cannot be observed from the outside. Such an assumption of information asymmetry also distinguishes our problem setting from the one considered in \cite{hibbard2019unpredictable}, where the ego agent and the adversary share the same partial observability and observation kernel. On the other hand, the transition kernel \eqref{equ:STK} and the nominal objective \eqref{equ:Reward} may be public knowledge, thus known by both the ego agent and the adversary.

Although not being able to observe $u_t$, based on the knowledge of \eqref{equ:STK} and \eqref{equ:Reward}, the adversary can infer the ego agent's actions $u_t$ through solving its nominal decision-making problem \eqref{equ:MDP}.

Once $\pi^*$ is obtained, the evolution of the state $x_t$ obeys the following Markov chain induced from \eqref{equ:STK} and $\pi^*$ in the eyes of the adversary:
\begin{equation}\label{equ:SMC}
    p_a (x'|x) := \mathbb{P}(x_{t+1} = x'|x_t = x,u_t = \pi^*(x)),\,\, \forall\,t \in \mathbb{N}_0,
\end{equation}
defined for all $x,x' \in X$.

Based on \eqref{equ:SMC}, the adversary can infer the state $x_t$, more specifically, the posterior belief of $x_t$ with all available observations, $\xi_t = \{y_0,\cdots,y_t\}$, taken into account, according to the following recursive Bayesian inference formula \citep{chen2003bayesian}:
\begin{align}\label{equ:Bayesian_a}
    & o(x'|\xi_t) := \mathbb{P}(x_t=x'|y_0,\cdots,y_t) \nonumber \\[3pt]
    &= \frac{\mathbb{P}(x_t=x',y_t|y_0,\cdots,y_{t-1})}{\mathbb{P}(y_t|y_0,\cdots,y_{t-1})} \nonumber \\[3pt]
    &= \frac{\mathbb{P}(y_t|y_0,\cdots,y_{t-1},x_t=x')\mathbb{P}(x_t=x'|y_0,\cdots,y_{t-1})}{\sum_{x'' \in X} \mathbb{P}(x_t=x'', y_t|y_0,\cdots,y_{t-1})} \nonumber \\[3pt]
    &= \frac{\mathbb{P}(y_t|x_t=x')\sum_{x \in X}\mathbb{P}(x_t=x',x_{t-1}=x|y_0,\cdots,y_{t-1})}{\sum_{x'' \in X}\mathbb{P}(y_t|x_t=x'') \mathbb{P}(x_t=x''|y_0,\cdots,y_{t-1})} \nonumber \\[3pt]
    &= \frac{q(y_t|x')\sum_{x \in X}\mathbb{P}(x_t=x'|y_0,\cdots,y_{t-1},x_{t-1}=x)o(x|\xi_{t-1})}{\sum_{x'' \in X}q(y_t|x'')\sum_{x \in X}\mathbb{P}(x_t=x'',x_{t-1}=x|y_0,\cdots,y_{t-1})} \nonumber \\[3pt]
    &= \frac{q(y_t|x')\sum_{x \in X}p_a (x'|x)o(x|\xi_{t-1})}{\sum_{x'' \in X} q(y_t|x'')\sum_{x \in X}p_a (x''|x)o(x|\xi_{t-1})}.
\end{align}

For convenience, we define the recursive Bayesian inference operator associated with the transition kernel $p_a$, $\mathcal{B}_a: \Delta \times Y \to \Delta$, where $\Delta:= \{o \in [0,1]^{|X|} \,|\, \|o\|_{\ell^1} = 1 \}$ is the $(|X|-1)$-dimensional probability simplex, as follows:
\begin{equation}\label{equ:Bayesian_operator}
\big(\mathcal{B}_a(o,y)\big)(x') = \frac{q(y|x')\sum_{x \in X}p_a (x'|x)o(x)}{\sum_{x'' \in X} q(y|x'')\sum_{x \in X}p_a (x''|x)o(x)}.
\end{equation}

\subsection{Decision-making with anti-detection objective}

We further consider the situation where the ego agent is aware of the existence of such an adversary and another objective of its decision-making is to hide from the adversary's detection.

In particular, we make the following assumptions: 1) The adversary is unaware of this anti-detection objective of the ego agent, so the adversary infers $x_t$ still relying on \eqref{equ:SMC} and \eqref{equ:Bayesian_a}. 2) The information leaked to the adversary, $y_t$, is known by the ego agent. And, 3) the previous belief of the adversary, $o(x|\xi_{t-1})$, is also known by the ego agent.

The assumptions 1) and 2) represent many real-world situations. The assumption 3) is also reasonable: the ego agent can start with a sufficiently nice initial prior (satisfying the Cromwell's rule \citep{jackman2009bayesian}), which is not necessarily identical to that of the adversary, e.g., a uniform distribution over $X$; then, the ego agent can recursively update the posterior belief associated with its own initial prior using \eqref{equ:Bayesian_a}; by the Bernstein--von Mises theorem \citep{van2000asymptotic}, the posterior belief is effectively asymptotically independent of the prior -- the ego agent's posterior belief will asymptotically converge to that of the adversary, and thus, after a sufficient number of time steps, the assumption 3) is approximately satisfied.

Based on the above assumptions, the ego agent can predict the adversary's future inferences using \eqref{equ:Bayesian_a}.

The extent of detection by the adversary can be measured by
\begin{equation}\label{equ:Stage_penalty}
    C \big(x_t,o(\cdot|\xi_t)\big) := o(x_t|\xi_t) = \sum_{x \in X} o(x|\xi_t) \mathbb{I}_x(x_t),
\end{equation}
where $x_t$ is the true value of the state. Using this measure, it makes no difference to the ego agent whether the distribution $o(\cdot|\xi_t)$ on $X$ is ``flat'' or has ``peaks'' at some $x \in X$ as long as $x \neq x_t$. This measure distinguishes our problem formulation from others, e.g., entropy or covariance maximization.

The anti-detection objective of the ego agent can then be represented using the following penalty function:
\begin{align}\label{equ:Penalty}
   & \mathcal{C}\big(x_t,o(\cdot|\xi_t)\big) = \\
   & \mathbb{E} \Big\{\sum_{\tau=0}^{\infty} \lambda^{\tau} C\big(x_{\tau|t},o(\cdot|\xi_{\tau|t})\big) \big| x_{0|t} = x_t, o(\cdot|\xi_{0|t}) = o(\cdot|\xi_t) \Big\}. \nonumber
\end{align}

{\it Remark 1:} By notation $o(\cdot|\xi_t)$, we mean a specific sequence of observations $\xi_t = \{y_0,\cdots,y_t\}$. Based on \eqref{equ:Bayesian_a} and \eqref{equ:Stage_penalty}, we have the following two observations: 1) The dynamics of $o(\cdot|\xi_t)$ have the Markov property, i.e., $o(\cdot|\xi_{t+1})$ is conditionally independent of $o(\cdot|\xi_{t-k})$, $k \in \mathbb{N}$, given $o(\cdot|\xi_t)$. In particular, $o(\cdot|\xi_{t+1})$ is uniquely determined by $o(\cdot|\xi_t)$ and $y_{t+1}$. 2) In general, it is possible that $o(\cdot|\xi_t,y_{t+1}^1) = o(\cdot|\xi_t,y_{t+1}^2)$ for $y_{t+1}^1 \neq y_{t+1}^2$. However, the penalty only depends on the value of $o(\cdot|\xi_t,y_{t+1}^1) = o(\cdot|\xi_t,y_{t+1}^2)$ rather than on which observation $y_{t+1}^1$ or $y_{t+1}^2$ is realized. Together with 1), it can be immediately seen that, on the one hand, the values of the posterior beliefs $o(\cdot|\xi_{\tau|t})$ have influences on the value of \eqref{equ:Penalty}, while on the other hand, the observations $y_{\tau|t}$ are just intermediate variables to propagate the posterior beliefs. To represent this fact and also simplify the notations, we use $o_t$ to denote the posterior belief at time step $t$, which can also be viewed as the equivalence class of $o(\cdot|\xi_t)$ with the equivalence relation defined by $o(\cdot|\xi_t^1) \sim o(\cdot|\xi_t^2)$ if $o(\cdot|\xi_t^1) = o(\cdot|\xi_t^2)$ where $\xi_t^1$ and $\xi_t^2$ denote any two different observation sequences.

Using the notation $o_t$, we formulate the decision-making problem of the ego agent when solely pursuing the anti-detection objective as:
\begin{equation}\label{equ:AD_MDP}
   \min_{\pi_a}\quad \mathcal{C}^{\pi_a}\big(x_t,o_t\big),
\end{equation}
where $\pi_a: X \times \Delta \to U$ is a policy such that $u_{\tau|t} = \pi_a(x_{\tau|t},o_{\tau|t})$ for all $\tau \in \mathbb{N}_0$, and
\begin{align}
& \mathcal{C}\big(x_t,o_t\big) = \mathbb{E} \Big\{\sum_{\tau=0}^{\infty} \lambda^{\tau} C\big(x_{\tau|t},o_{\tau|t}\big) \big| x_{0|t} = x_t, o_{0|t} = o_t\Big\}, \nonumber \\
& C\big(x_t,o_t\big) = \sum_{x \in X} o_t(x) \mathbb{I}_x(x_t),
\end{align}
subject to a transition kernel representing the dynamics of $(x_t,o_t,u_t) \to (x_{t+1},o_{t+1})$.

Note that $o_t$ is a distribution on a finite set represented using a vector in $\Delta$, in particular, $o_t$ takes continuous values. Thus, the transition kernel for $(x_t,o_t)$ cannot be represented using a time-invariant transition matrix as can be done for \eqref{equ:STK}. Fortunately, it admits a closed-form expression using the recursive Bayesian inference operator $\mathcal{B}_a$ defined in \eqref{equ:Bayesian_operator} as follows:
\begin{align}\label{equ:OTK}
   & r(x',o'|x,o,u) \nonumber \\[2pt]
   &:= \mathbb{P} \big(x_{t+1} = x', o_{t+1} = o' \big| x_t = x, o_t = o, u_t = u\big) \nonumber \\[4pt]
   &= \sum_{y \in Y} \mathbb{P} \big(x_{t+1} = x', o_{t+1} = o', \nonumber \\[-8pt]
   &\quad\quad\quad\quad\quad\quad y_{t+1} = y \big|  x_t = x, o_t = o, u_t = u\big) \nonumber \\[4pt]
   &= \sum_{y \in Y} \Big(\mathbb{P} \big(o_{t+1} = o' \big| o_t = o, y_{t+1} = y\big) \cdot \nonumber \\[-2pt]
   &\quad \cdot \mathbb{P} \big(y_{t+1} = y \big| x_{t+1} = x' \big) \mathbb{P} \big(x_{t+1} = x' \big| x_t = x, u_t = u \big)\Big) \nonumber \\[4pt]
   &= \sum_{y \in Y} \Big(\mathbb{I}_{\mathcal{B}_a(o, y)}(o')\, q(y|x')\, p(x'|x,u)\Big).
\end{align}

{\it Lemma 1:} Given a 3-tuple $(x,o,u) \in X \times \Delta \times U$, the $r(\cdot,\cdot|x,o,u)$ defined by \eqref{equ:OTK} is either ill-defined or a discrete probability measure on $X \times \Delta$ with finite support.

{\it Proof:} Note that the recursive Bayesian inference operator $\mathcal{B}_a$ defined by \eqref{equ:Bayesian_operator} is ill-defined for $(o,y) \in \Delta \times Y$ such that $q(y|x')\sum_{x \in X}p_a (x'|x)o(x) = 0$ for all $x' \in X$ (causing $\frac{0}{0}$). Consequently, the $r(\cdot,\cdot|x,o,u)$ defined by \eqref{equ:OTK} is ill-defined for $(x,o,u) \in X \times \Delta \times U$ such that there exists $y \in Y$ with $\sum_{x' \in X} q(y|x')\, p(x'|x,u)>0$ and $\mathcal{B}_a$ being ill-defined for $(o,y) \in \Delta \times Y$.

When $r(\cdot,\cdot|x,o,u)$ is well-defined, for each $x' \in X$, there are at most $|Y|$ different points $o' \in \Delta$ with $r(x',o'|x,o,u)$ being non-zero. This fact follows from the term $\mathbb{I}_{\mathcal{B}_a(o, y)}(o')$ in \eqref{equ:OTK} and the fact that $\mathcal{B}_a: \Delta \times Y \to \Delta$ is a deterministic operator. Thus, due to the finiteness of $X \times Y$, $r(\cdot,\cdot|x,o,u)$ has finite support on $X \times \Delta$. It is also clear from \eqref{equ:OTK} that $r(\cdot,\cdot|x,o,u)$ is non-negative. Then, it remains to show that $\int_{X \times \Delta} \text{d}r(\cdot,\cdot|x,o,u) = \sum_{x' \in X} \sum_{o' \in \Delta} r(x',o'|x,o,u) = 1$.\footnote{Most generally, the sum $\sum_{o' \in \Delta}$ can be defined as $\sum_{o' \in \Delta} := \sup \big\{\sum_{o' \in \Delta'} |\, \Delta' \subset \Delta \text{ is finite}\, \big\}$.} This holds since
\begin{align}
   & \sum_{x' \in X} \sum_{o' \in \Delta} r(x',o'|x,o,u) \nonumber \\
   =&\, \sum_{x' \in X} \sum_{o' \in \Delta} \sum_{y \in Y} \Big(\mathbb{I}_{\mathcal{B}_a(o, y)}(o')\, q(y|x')\, p(x'|x,u)\Big) \label{equ:sum_1} \\
   =&\, \sum_{x' \in X} \sum_{y \in Y} \Big(\sum_{o' \in \Delta} \mathbb{I}_{\mathcal{B}_a(o, y)}(o')\, q(y|x')\, p(x'|x,u)\Big) \label{equ:sum_2} \\
   =&\, \sum_{x' \in X} \sum_{y \in Y} q(y|x')\, p(x'|x,u) = \sum_{x' \in X} p(x'|x,u) = 1, \nonumber
\end{align}
where Tonelli's theorem enables us to switch the order of the summation and obtain \eqref{equ:sum_2} from \eqref{equ:sum_1}. Therefore, $r(\cdot,\cdot|x,o,u)$ is a discrete probability measure on $X \times \Delta$ with finite support. $\blacksquare$

{\it Remark 2:} The fact that the recursive Bayesian inference operator $\mathcal{B}_a$ is ill-defined for pairs $(o,y) \in \Delta \times Y$ with $\sum_{x' \in X} q(y|x')\sum_{x \in X}p_a (x'|x)o(x) = 0$ is not a problem for traditional applications of Bayesian estimation, e.g., in the settings of hidden Markov models and partially observable MDPs, because such $(o,y)$ pairs can {\it almost never}, i.e., with probability $0$, occur. However, in our setting, due to the possibility for the ego agent's action $u_t = u$ to be different from the adversary's prediction $u_t = \pi^*(x_t)$, such $(o_t,y_{t+1}) = (o,y)$ may occur. Physically, this represents a case where the adversary observes an observation $y_{t+1} = y$ that can almost never occur if the ego agent applies the policy $\pi^*$. This may shock the adversary and let it become aware of the ego agent's anti-detection behavior, which may be undesirable. Therefore, the occurrence of such cases should be avoided. The way to avoid it will be discussed in detail in Section~\ref{sec:3}.

To account for both the nominal objective \eqref{equ:Reward} and the objective of anti-detection \eqref{equ:Penalty}, two different strategies can be pursued: The first strategy is to treat one of \eqref{equ:Reward} and \eqref{equ:Penalty} as the objective function to optimize and impose a constraint representing the requirement that the value of the other has to be higher/lower than a specified threshold \citep{savas2018entropy,hibbard2019unpredictable}. In this paper, we pursue another strategy, i.e., to optimize a convex combination of \eqref{equ:Reward} and \eqref{equ:Penalty}. In particular, we consider the following decision-making process:
\begin{subequations}\label{equ:MDP_c}
\begin{align}
   \max_{\pi_a}\quad & w_n\, \mathcal{R}^{\pi_a}(x_t) - w_a\, \mathcal{C}^{\pi_a}(x_t,o_t), \\
   \text{s.t.}\quad & r(x',o'|x,o,u),
\end{align}
\end{subequations}
where $w_n,w_a \in [0,\infty)$ are weighting factors that balance the considerations of nominal and anti-detection objectives.

A side observation is that when $w_a<0$, \eqref{equ:MDP_c} is increasing the probability of being detected instead of decreasing it. Such a formulation may be useful in the context of information collection in partially observable environments.

\section{Solution approach based on value iteration}\label{sec:3}

In this section, we describe a value iteration (VI) approach to solve the problem \eqref{equ:MDP_c}.

Note that, given a pair $(x,o)$, $u$ must take its value in $U_a(x,o) = U \setminus U_p(x,o)$ for the well-definedness of $r(\cdot,\cdot|x,o,u)$, where the set of prohibited actions $U_p(x,o)$ is defined as follows (see proof of Lemma~1):
\begin{align}\label{equ:prohibited_set}
& U_p(x,o) := \nonumber \\[4pt]
& \Big\{u \in U \,|\, \exists\, y \in Y \text{ s.t.} \sum_{x' \in X} q(y|x')\, p(x'|x,u) > 0 \nonumber \\[-2pt]
&\quad\,\, \text{and} \sum_{x'' \in X} q(y|x'') \big(\sum_{x' \in X} p_a (x''|x') o(x')\big) = 0 \Big\}.
\end{align}

We are now ready to present the main result of this section:

{\it Theorem 1:} Assume that $U_a(x,o) \neq \emptyset$ for all $(x,o) \in X \times \Delta$.\footnote{Supported by Lemma~2 in Appendix.} Then, the solution to \eqref{equ:MDP_c} over the set of admissible policies, $\Pi_a$, satisfying $\pi_a(x,o) \in U_a(x,o)$ for all $(x,o) \in X \times \Delta$, exists and satisfies the following Bellman equation:
\begin{align}\label{equ:Bellman_c}
    & V_a(x_t,o_t) = \max_{u \in U_a(x_t,o_t)} \Big\{ \big(w_n\, R(x_t,u) -w_a\, C(x_t,o_t)\big) \nonumber \\
    &\quad + \lambda\, \mathbb{E} \big\{V_a(x_{1|t},o_{1|t}) \big| x_{0|t} = x_t, o_{0|t} = o_t, u_{0|t} = u \big\} \Big\}, \nonumber \\
    & = \max_{u \in U_a(x_t,o_t)} \Big\{ \big(w_n\, R(x_t,u) -w_a\, C(x_t,o_t)\big) \nonumber \\
    &\quad + \lambda \sum_{x' \in X}\sum_{o' \in \Delta} \big( r(x',o'|x_t,o_t,u) V_a(x',o')\big)\Big\},
\end{align}
where $V_a(x,o):= \max_{\pi_a \in \Pi_a}\, w_n\, \mathcal{R}^{\pi_a}(x) - w_a\, \mathcal{C}^{\pi_a}(x,o)$ can be determined by the following value iteration:
\begin{align}\label{equ:VI_c}
    & V_a^{k+1}(x,o) = \max_{u \in U_a(x,o)} \Big\{ \big(w_n\, R(x,u) -w_a\, C(x,o)\big) \nonumber \\
    &\quad + \lambda \sum_{x' \in X}\sum_{o' \in \Delta} \big( r(x',o'|x,o,u) V_a^k (x',o')\big)\Big\},
\end{align}
as $k \to \infty$. Note that the sum of $r(x',o'|x,o,u) V_a (x',o')$ over $X \times \Delta$ is finite by the fact that $r(\cdot,\cdot|x,o,u)$ has finite support.

Once $V_a(x,o)$ is obtained, the optimal policy for the ego agent under weighted nominal and anti-detection decision-making objectives, $\pi_a^*$, can be obtained by
\begin{align}\label{equ:C_P}
    & \pi_a^*(x,o) = \argmax_{u \in U_a(x,o)} \Big\{ \big(w_n\, R(x,u) -w_a\, C(x,o)\big) \nonumber \\
    &\quad + \lambda \sum_{x' \in X}\sum_{o' \in \Delta} \big( r(x',o'|x,o,u) V_a(x',o')\big)\Big\}.
\end{align}

{\it Proof:} Note that the following properties hold for \eqref{equ:MDP_c}:

1) For each state $(x,o) \in X \times \Delta$, the set $U_a(x,o)$ of admissible actions (non-empty by assumption) is finite, thus, compact.

2) Since the reward $R: X \times U \to \mathbb{R}$ is defined on a finite space and the penalty $C: X \times \Delta \to \mathbb{R}$ takes values in $[0,1]$, the $w_n\, R -w_a\, C$ is bounded on the set of admissible state-action pairs $\big\{\big((x,o),u\big) \big| (x,o) \in X \times \Delta, u \in U_a(x,o)\big\}$. Moreover, for each $(x,o) \in X \times \Delta$, $w_n\, R(x,\cdot) -w_a\, C(x,o)$ is a continuous function of $u \in U_a(x,o)$, which is trivial since $U_a(x,o)$ is finite/discrete.

3) For each $(x,o) \in X \times \Delta$ and $v \in L^{\infty}(X \times \Delta)$\footnote{$L^{\infty}(X \times \Delta)$ denotes the Banach space of real-valued bounded measurable functions on $X \times \Delta$ with the sup-norm $\|v\|:= \sup_{(x,o)} |v(x,o)|$.}, $\int_{X \times \Delta} v(x',o')\, \text{d}r(x',o'|x,o,\cdot) = \sum_{x' \in X} \sum_{o' \in \Delta}$ $\big(v(x',o')\, r(x',o'|x,o,\cdot)\big)$ is a continuous function of $u \in U_a(x,o)$, which is trivial, again, due to the finiteness of $U_a(x,o)$.

Then, the statements of Theorem~1 follow from Theorem~2.2 and Remark~2.3 of \cite{hernandez2012adaptive}. $\blacksquare$

Since $o \in \Delta$ takes continuous values, to numerically operate the value iteration \eqref{equ:VI_c} requires functional approximation techniques. In practice, we discretize the space $\Delta$, update the values of $V_a^{k+1}(x,o)$ on the grid points, during which we obtain the values of $V_a^{k}(x',o')$ through interpolation.

Note also that we need not characterize the set \eqref{equ:prohibited_set} beforehand, but can identify the events $u\, ^{\in}\!/\!_{\notin} \,U_p(x,o)$ when computing $r(x',o'|x,o,u)$ (defined in \eqref{equ:OTK}) in the course of evaluating \eqref{equ:VI_c} or \eqref{equ:C_P}.

\section{Approximate solution approach based on receding-horizon optimization}\label{sec:4}

The value iteration approach \eqref{equ:VI_c} to solving \eqref{equ:MDP_c} suffers from the ``curse of dimensionality." In particular, similarly to the situation in partially observable MDPs, as the number of states $x \in X$ increases, the dimension of the posterior belief vector $o \in \Delta \subset [0,1]^{|X|}$ increases. Consequently, the value iteration becomes computationally challenging when the finite state space $X$ becomes larger.

To handle problems of larger size, we propose a receding-horizon approximation to \eqref{equ:MDP_c} as follows:
\begin{subequations}\label{equ:MDP_c_approxi_1}
\begin{align}
   \max_{\mu_{0:N-1},\, \pi}\quad & w_n\,(\mathcal{R}_1 + \mathcal{R}_2) - (w_a + w_a')\, \mathcal{R}_3,  \\
   \text{s.t.}\quad &  r(x',o'|x,o,u),
\end{align}
\end{subequations}
where
\begin{align}\label{equ:MDP_c_approxi_2}
   & \mathcal{R}_1 = \mathbb{E} \Big\{\sum_{\tau=0}^{N-1} \lambda^{\tau} R(x_{\tau|t},u_{\tau|t}) \big|x_{0|t} = x_t, \nonumber \\[-4pt]
   &\quad\quad\quad\quad\quad\quad\quad\quad\quad\quad \{u_{\tau|t}\}_{\tau=0}^{N-1} = \mu_{0:N-1} \Big\}, \nonumber \\
   & \mathcal{R}_2 = \mathbb{E} \Big\{\sum_{\tau=N}^{\infty} \lambda^{\tau} R(x_{\tau|t},u_{\tau|t}) \big| x_{0|t} = x_t, \nonumber \\[-4pt]
   &\quad\quad\quad\,\,\,\, \{u_{\tau|t}\}_{\tau=0}^{N-1} = \mu_{0:N-1}, u_{\tau|t} = \pi(x_{\tau|t}),\, \forall\, \tau \ge N \Big\}, \nonumber \\
   & \mathcal{R}_3 = \mathbb{E} \Big\{\sum_{\tau=1}^{N-1} \lambda^{\tau} C(x_{\tau|t},o_{\tau|t}) \big| x_{0|t} = x_t, \nonumber \\[-4pt]
   &\quad\quad\quad\quad\quad\quad o_{0|t} = o_t, \{u_{\tau|t}\}_{\tau=0}^{N-1} = \mu_{0:N-1} \Big\}.
\end{align}
Note that in defining $\mathcal{R}_3$ we do not include the term $\lambda^{\tau} C(x_{\tau|t},o_{\tau|t})$ associated with $\tau = 0$ as it is a constant and does not change the maximizers.

The approximation \eqref{equ:MDP_c_approxi_1} is to use open-loop actions $\mu_{0:N-1} = \{\mu_0, \mu_1, \cdots, \mu_{N-1}\}$ to control the ego agent for the first $N$ steps over the horizon, then switch to closed-loop policy $\pi: X \to U$ for the remaining horizon. In particular, $w_n \mathcal{R}_1 - w_a \mathcal{R}_3$ represents the expected value of the cumulative reward over the first $N$ steps, $\sum_{\tau=0}^{N-1} \lambda^{\tau} \big(w_n R(x_{\tau|t},u_{\tau|t}) -w_a C(x_{\tau|t},o_{\tau|t})\big)$ (up to a constant); $\mathcal{R}_2$ represents the expected value of the cumulative reward associated with the nominal objective over the remaining horizon; and finally, $w_a' \mathcal{R}_3$ represents an estimate of the cumulative penalty associated with the anti-detection objective over the remaining horizon based on that over the first $N$ steps, where $w_a'$ is a tuning parameter suggested to be picked in the range $\big[0,\frac{\lambda^N}{1-\lambda^N} w_a\big]$.

After solving \eqref{equ:MDP_c_approxi_1}, the ego agent applies $\mu_0$ as the action for time step $t$, i.e., $u_t = \mu_0$. Once the state has been updated from $x_t$ to $x_{t+1}$, the ego agent repeats the same procedure at the next time step $t+1$.

{\it Theorem 2:} Problem \eqref{equ:MDP_c_approxi_1} can be equivalently\footnote{In terms of maintaining the maximizers $\mu_{0:N-1}$ to be the same as those for \eqref{equ:MDP_c_approxi_1}.} simplified to
\begin{subequations}\label{equ:MDP_c_approxi_3}
\begin{align}
   \max_{\mu_{0:N-1}}\quad & w_n\,(\mathcal{R}_1 + \mathcal{R}_2^*) - (w_a + w_a')\, \mathcal{R}_3, \\
   \text{s.t.}\quad & r(x',o'|x,o,u),
\end{align}
\end{subequations}
with
\begin{align}\label{equ:MDP_c_approxi_4}
   & \mathcal{R}_2^* = \lambda^{N} \sum_{x \in X} \Big( V(x) \cdot \\[-2pt]
   &\quad\quad \cdot \mathbb{P} \big( x_{N|t} = x\big|x_{0|t} = x_t, \{u_{\tau|t}\}_{\tau=0}^{N-1} = \mu_{0:N-1}\big)\Big), \nonumber
\end{align}
where $V(x)$ is the value function defined in \eqref{equ:Bellman} and has been computed when solving \eqref{equ:MDP}.

{\it Proof:} Note that 1) $\mathcal{R}_1$ and $\mathcal{R}_3$ are independent of $\pi$, and 2) using the Markov property of the dynamics of $x$, $\mathcal{R}_2$ can be written as
\begin{align}\label{equ:Approxi_R2}
   & \mathcal{R}_2 = \sum_{x \in X} \bigg( \mathbb{E} \Big\{\sum_{\tau=N}^{\infty} \lambda^{\tau} R(x_{\tau|t},u_{\tau|t}) \big| x_{N|t} = x, \nonumber \\[-2pt]
   &\quad\quad u_{\tau|t} = \pi(x_{\tau|t}),\, \forall\, \tau \ge N \Big\} \mathbb{P} \Big( x_{N|t} = x \big| x_{0|t} = x_t, \nonumber \\[-2pt]
   &\quad\quad \{u_{\tau|t}\}_{\tau=0}^{N-1} = \mu_{0:N-1}\Big)\bigg) \nonumber \\
   &= \lambda^{N} \sum_{x \in X} \bigg( \mathbb{E} \Big\{\sum_{\tau=0}^{\infty} \lambda^{\tau} R(x_{\tau|t},u_{\tau|t}) \,\big|\, x_{0|t} = x, \nonumber \\[-2pt]
   &\quad\quad u_{\tau|t} = \pi(x_{\tau|t}),\, \forall\, \tau \ge 0 \Big\} \mathbb{P} \Big( x_{N|t} = x \big| x_{0|t} = x_t, \nonumber \\[-2pt]
   &\quad\quad \{u_{\tau|t}\}_{\tau=0}^{N-1} = \mu_{0:N-1}\Big)\bigg) \\
   &= \lambda^{N} \sum_{x \in X} \Big( \mathcal{R}^{\pi}(x) \cdot \nonumber \\[-2pt]
   &\quad \cdot \mathbb{P} \big( x_{N|t} = x \big| x_{0|t} = x_t, \{u_{\tau|t}\}_{\tau=0}^{N-1} = \mu_{0:N-1}\big)\Big). \nonumber
\end{align}
Then, based on Bellman's principle of optimality \citep{Bellman:2003:DP:862270}, the optimal policy $\pi$ to problem \eqref{equ:MDP_c_approxi_1} agrees with the optimal solution to \eqref{equ:MDP}, and the statement of Theorem~2 follows. $\blacksquare$

Problem \eqref{equ:MDP_c_approxi_3} is a finite-horizon ($N$-step look-ahead) open-loop decision-making problem with a finite decision set, and can be solved online using a tree-search method. Note that at each prediction step, only a finite number of values for the pair $(x',o')$ needs to be accounted for, due to the fact that, given $(x,o)$, $r(\cdot,\cdot|x,o,u)$ has finite support for each $u$. In particular, for any action sequence $\mu_{0:N-1}$, $\mathcal{R}_1$, $\mathcal{R}_2^*$, and $\mathcal{R}_3$ can be evaluated according to \eqref{equ:MDP_c_approxi_4} and the following \eqref{equ:R1_eval} \eqref{equ:R3_eval},
\begin{align}
   \mathcal{R}_1 &= \sum_{\tau=0}^{N-1} \lambda^{\tau} \Big(\sum_{x \in X} R(x,\mu_{\tau}) \mathbb{P} \big(x_{\tau|t} = x \big|x_{0|t} = x_t, \nonumber \\[-4pt]
   &\quad\quad\quad\quad\quad\quad\quad\quad \{u_{k|t}\}_{k=0}^{\tau-1} = \mu_{0:\tau-1} \big)\Big), \label{equ:R1_eval} \\
   \mathcal{R}_3 &= \sum_{\tau=1}^{N-1} \lambda^{\tau} \Big(\sum_{x \in X} \sum_{o \in \Delta} C(x,o) \mathbb{P} \big((x_{\tau|t},o_{\tau|t}) = (x,o) \big| \nonumber \\[-4pt]
   &\quad\quad x_{0|t} = x_t, o_{0|t} = o_t, \{u_{k|t}\}_{k=0}^{\tau-1} = \mu_{0:\tau-1} \big)\Big), \label{equ:R3_eval}
\end{align}
where for every $\tau = 1, \cdots, N$,
\begin{align}
   & \mathbb{P} \big((x_{\tau|t},o_{\tau|t}) = (x',o') \big| x_{0|t} = x_t, o_{0|t} = o_t, \nonumber \\[2pt]
   &\quad\quad \{u_{k|t}\}_{k=0}^{\tau-1} = \mu_{0:\tau-1} \big) = \nonumber \\[4pt]
   & \sum_{x \in X}\! \sum_{o \in \Delta} \mathbb{P} \big((x_{\tau|t},o_{\tau|t}) = (x',o'), (x_{\tau-1|t},o_{\tau-1|t}) = \nonumber \\[-2pt]
   &\quad (x,o) \big| x_{0|t} = x_t, o_{0|t} = o_t, \{u_{k|t}\}_{k=0}^{\tau-1} = \mu_{0:\tau-1} \big) = \nonumber \\[6pt]
   &\sum_{x \in X}\! \sum_{o \in \Delta} r(x',o'|x,o,\mu_{\tau-1}) \mathbb{P} \big((x_{\tau-1|t},o_{\tau-1|t}) = \nonumber \\[-2pt]
   &\quad (x,o) \big| x_{0|t} = x_t, o_{0|t} = o_t, \{u_{k|t}\}_{k=0}^{\tau-2} = \mu_{0:\tau-2} \big),
\end{align}
computed recursively, and
\begin{align}
& \mathbb{P} \big(x_{\tau|t} = x \big| x_{0|t} = x_t, \{u_{k|t}\}_{k=0}^{\tau-1} = \mu_{0:\tau-1} \big) \nonumber \\[6pt]
&= \sum_{o \in \Delta} \mathbb{P} \big((x_{\tau|t},o_{\tau|t}) = (x,o) \big| x_{0|t} = x_t, o_{0|t} = o_t, \nonumber \\[-4pt]
&\quad\quad\quad\quad\quad\quad \{u_{k|t}\}_{k=0}^{\tau-1} = \mu_{0:\tau-1} \big).
\end{align}

\section{Examples}\label{sec:5}

\subsection{Example 1}

We consider an MDP model with $X = \{1,2,3\}$ and $U = \{1,2\}$, as follows:
\begin{equation}
p^1 = \begin{bmatrix} 0.8 & 0.1 & 0.1 \\
        0.1 & 0.8 & 0.1 \\
        0.1 & 0.1 & 0.8 \end{bmatrix}, \quad
p^2 = \begin{bmatrix} 0.1 & 0.1 & 0.8 \\
        0.8 & 0.1 & 0.1 \\
        0.1 & 0.8 & 0.1 \end{bmatrix},
\end{equation}
where $p_{ij}^k = p(i|j,k)$; the nominal reward function \eqref{equ:Reward} is with
\begin{equation}
R(x,u) = \mathbb{I}_{1}(x) + 0.8\, \mathbb{I}_{2}(x),\quad \lambda = 0.95;
\end{equation}
and the observation kernel of the adversary is
\begin{equation}
q = \begin{bmatrix} 0.7 & 0.1 & 0.05 \\
    0.15 & 0.45 & 0.05 \\
    0.15 & 0.45 & 0.9 \end{bmatrix},
\end{equation}
where $q_{ij} = q(i|j)$.

\begin{figure}[ht]
\begin{center}
\begin{picture}(300.0, 100.0)
\put(  0,  -12.5){\epsfig{file=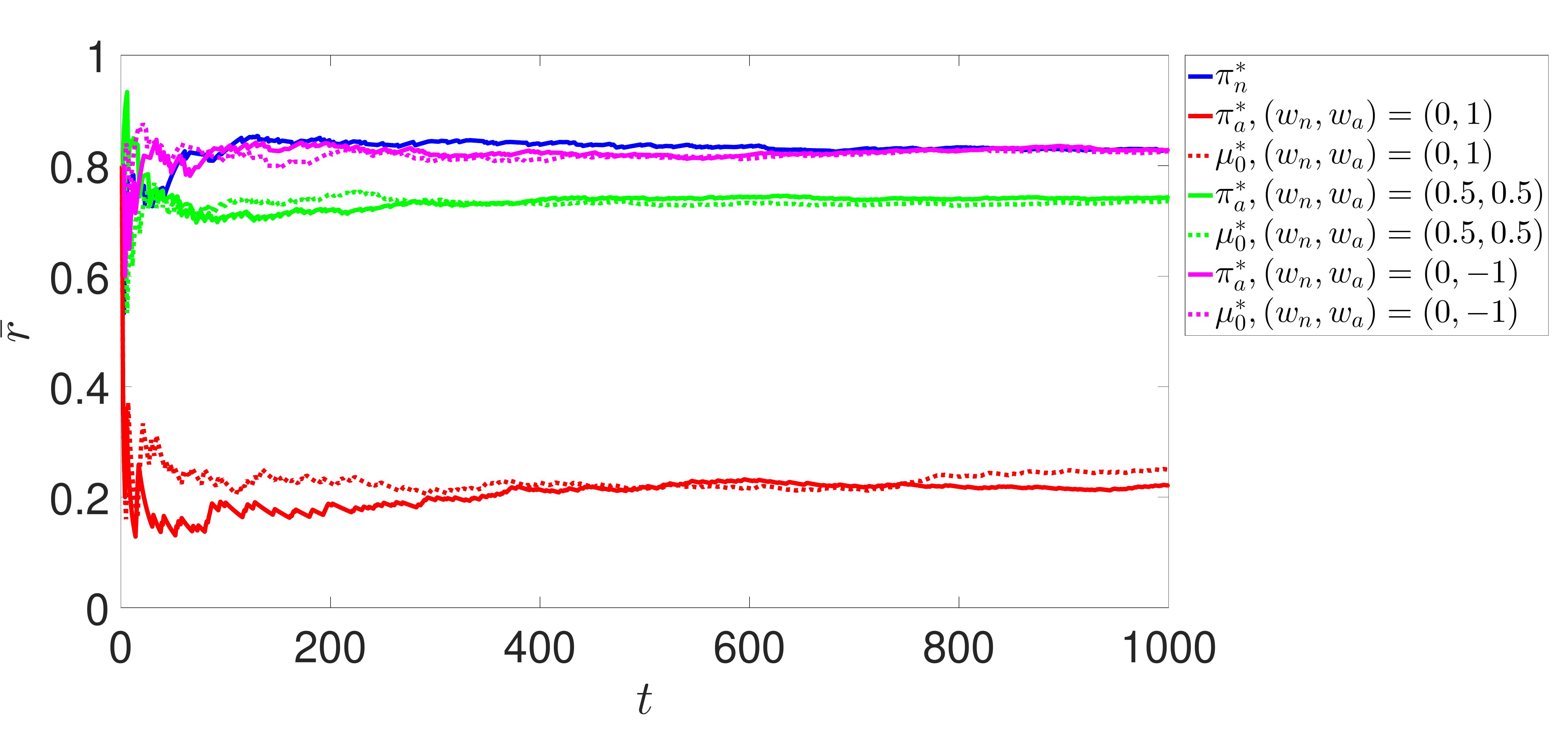,width = 1 \linewidth}}
\end{picture}
\end{center}
      \caption{Example 1: Time history of average reward.}
      \label{fig: EX1_R}
\end{figure}

\begin{figure}[ht]
\begin{center}
\begin{picture}(300.0, 100.0)
\put(  0,  -12.5){\epsfig{file=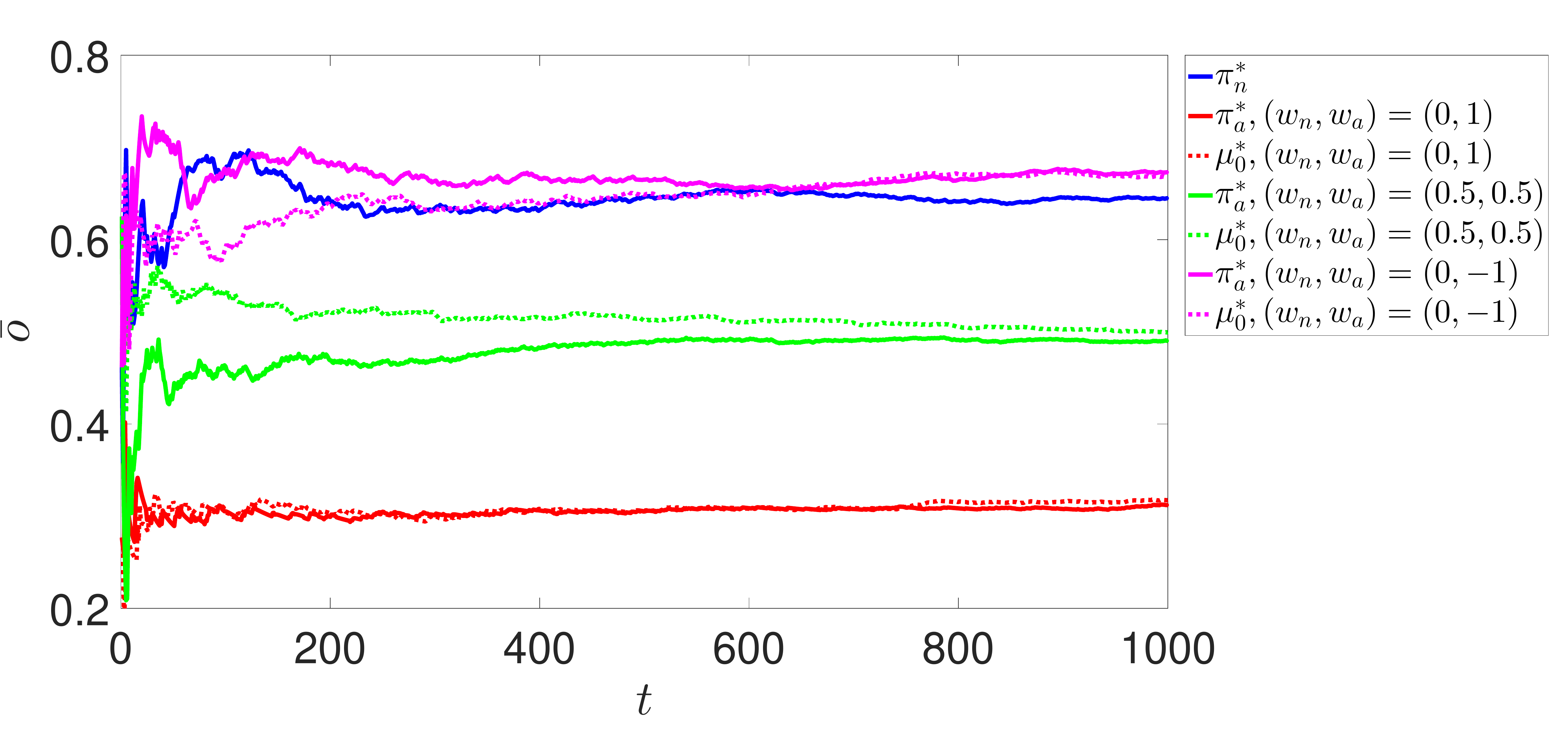,width = 1 \linewidth}}
\end{picture}
\end{center}
      \caption{Example 1: Time history of average extent of detection.}
      \label{fig: EX1_O}
\end{figure}

We use, respectively, the policy $\pi^*$ obtained by solving \eqref{equ:MDP}, the policy $\pi_a^*$ obtained by solving \eqref{equ:MDP_c}, and the online determined actions $\mu_0^*$ from \eqref{equ:MDP_c_approxi_3} with $N=3$ and $w_a' = 0$ to control the ego agent, and we run closed-loop simulations for different combinations of $w_n$ and $w_a$. We plot the average reward achieved by the ego agent $\bar{r}(t) = \frac{1}{t+1} \sum_{k=0}^t R(x_k,u_k)$ and the average extent of detection by the adversary $\bar{o}(t) = \frac{1}{t+1} \sum_{k=0}^t o(x_k|\xi_k)$ during the simulation for each case in Figs.~\ref{fig: EX1_R} and \ref{fig: EX1_O}, where the blue solid curves correspond to $\pi^*$, the red, green, magenta solid curves correspond to $\pi_a^*$ with $(w_n,w_a) = (0,1),\, (0.5,0.5),\, (0,-1)$, and the red, green, magenta dotted curves correspond to $\mu_0^*$ with $(w_n,w_a) = (0,1),\, (0.5,0.5),\, (0,-1)$.

It can be observed that for $(w_n,w_a) = (0,1)$, i.e., the ego agent only caring about anti-detection, the extent of detection measured by $\bar{o}(t)$ is maintained to be low using either the offline constructed policy $\pi_a^*$ or the online determined actions $\mu_0^*$, in particular, much lower as compared to the nominal policy $\pi^*$. However, the nominal reward $\bar{r}(t)$ is also very low, representing an unsatisfactory behavior in terms of executing the nominal mission.

For $(w_n,w_a) = (0.5,0.5)$, with a small sacrifice in the nominal reward $\bar{r}(t)$ (less than $0.1$), the ego agent can reduce the extent of detection $\bar{o}(t)$ by a considerable amount (more than $0.15$)\footnote{Note that the maximum achievable decrease in $\bar{o}(t)$ is about $0.3$.}.

For $(w_n,w_a) = (0,-1)$, i.e., the ego agent maximizing the extent of detection, the resulting $\bar{o}(t)$ is higher than that corresponding to the nominal policy $\pi^*$, while the nominal reward $\bar{r}(t)$ is close to that of $\pi^*$, representing the fact that, for most time instants, the action maximizing the extent of detection agrees with the one generated from the nominal policy $\pi^*$.

Furthermore, it can be observed that the receding-horizon solutions from \eqref{equ:MDP_c_approxi_3} approximate the policy $\pi_a^*$ satisfactorily, since their performance in terms of $\bar{r}(t)$ and $\bar{o}(t)$ are close.

\subsection{Example 2}

The second example we consider represents a path planning problem for an ego agent in a grid world, where, in addition to reaching a target location, the ego agent wants to hide its location from detection by a radar sensor.

\begin{figure}[ht]
\begin{center}
\begin{picture}(160.0, 128.0)
\put(  0,  -10){\epsfig{file=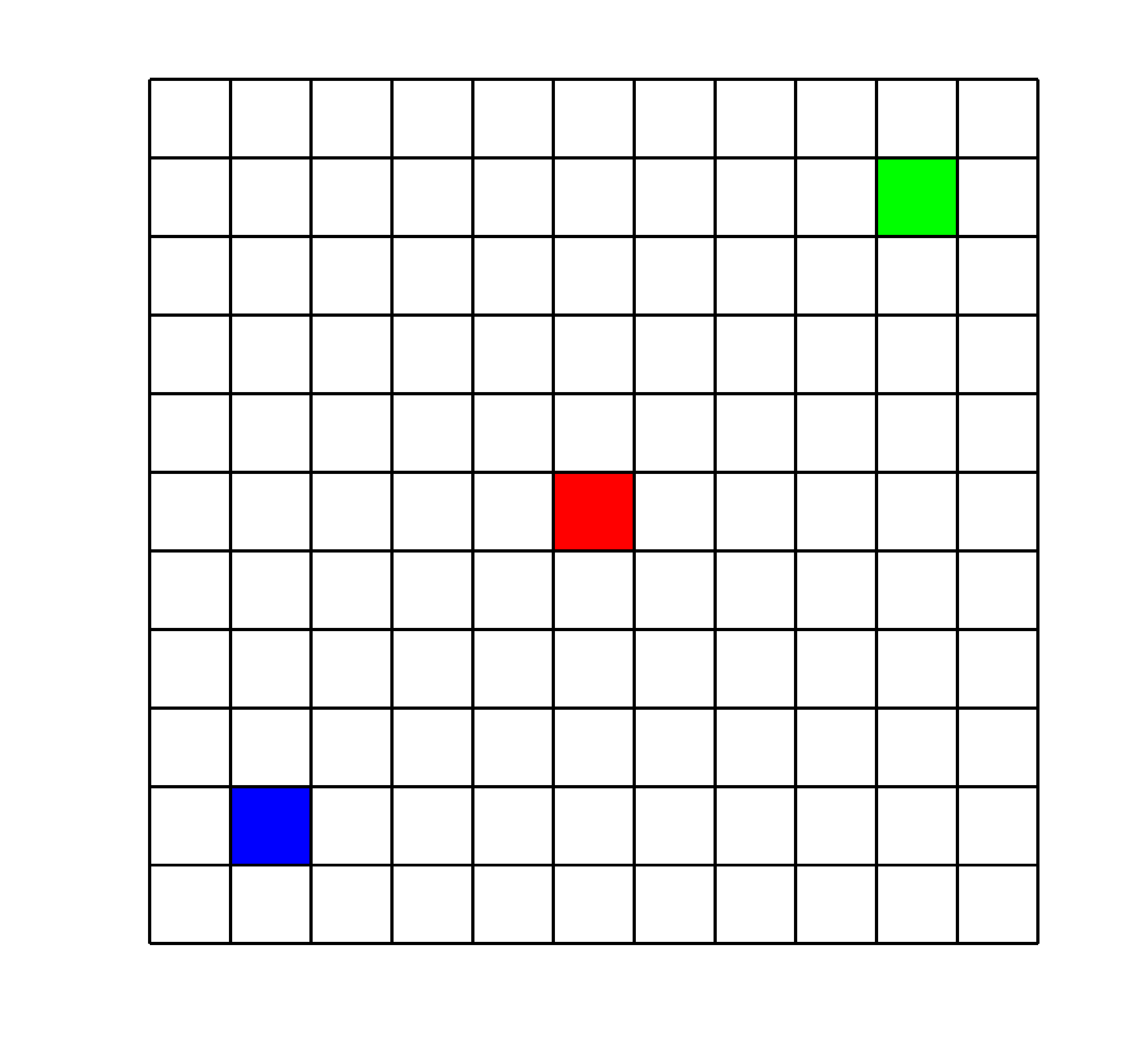,width = 0.64 \linewidth}}
\end{picture}
\end{center}
      \caption{Example 2: Map of the grid world. The blue, green, and red grid squares denote, respectively, the initial and target locations of the ego agent, and the location of the sensor.}
      \label{fig: EX2_map}
\end{figure}

The initial and target locations of the ego agent, and the location of the sensor are marked, respectively, by blue, green, and red on the map (Fig.~\ref{fig: EX2_map}). It is assumed that the sensor can measure the distance in $\ell^1$-norm between the ego agent and itself, where the measurement is corrupted by discretized Gaussian noise.

In this example, due to the large state space $X$ ($|X| = 121$), the computation load associated with the value iteration approach \eqref{equ:MDP_c} is heavy. Thus, we apply the receding-horizon optimization approach \eqref{equ:MDP_c_approxi_3} to solve this problem.

\begin{figure}[ht]
\begin{center}
\begin{picture}(160.0, 132.0)
\put(  0,  -10){\epsfig{file=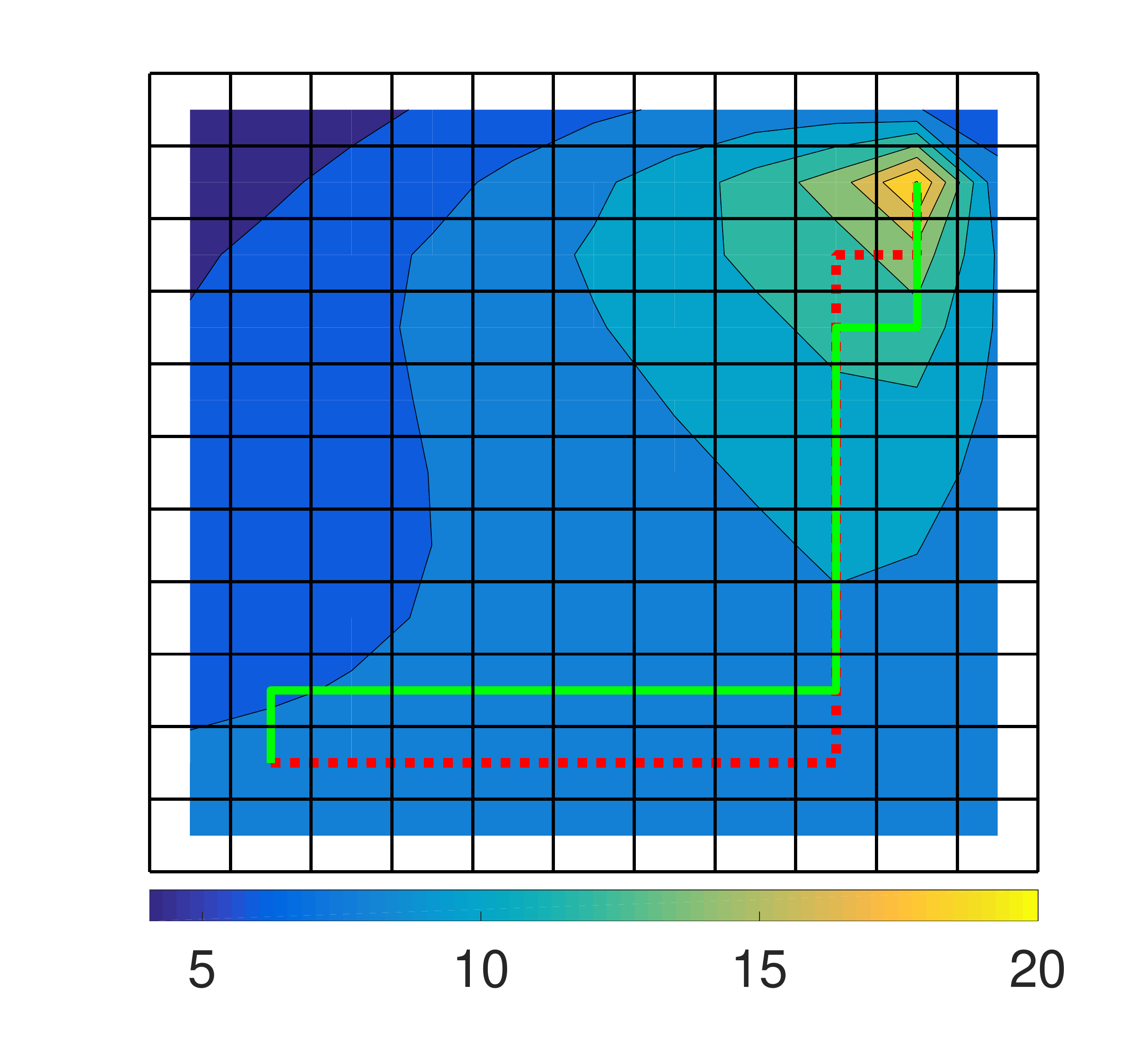,width = 0.64 \linewidth}}
\end{picture}
\end{center}
      \caption{Example 2: Contour of the value function $V(x)$, and the paths traversed by the ego agent when solely pursuing the nominal objective (in red dotted) and when pursuing both the nominal and the anti-detection objectives (in green solid).}
      \label{fig: EX2_path}
\end{figure}

\begin{figure}[ht]
\begin{center}
\begin{picture}(160.0, 135.0)
\put(  0,  -10){\epsfig{file=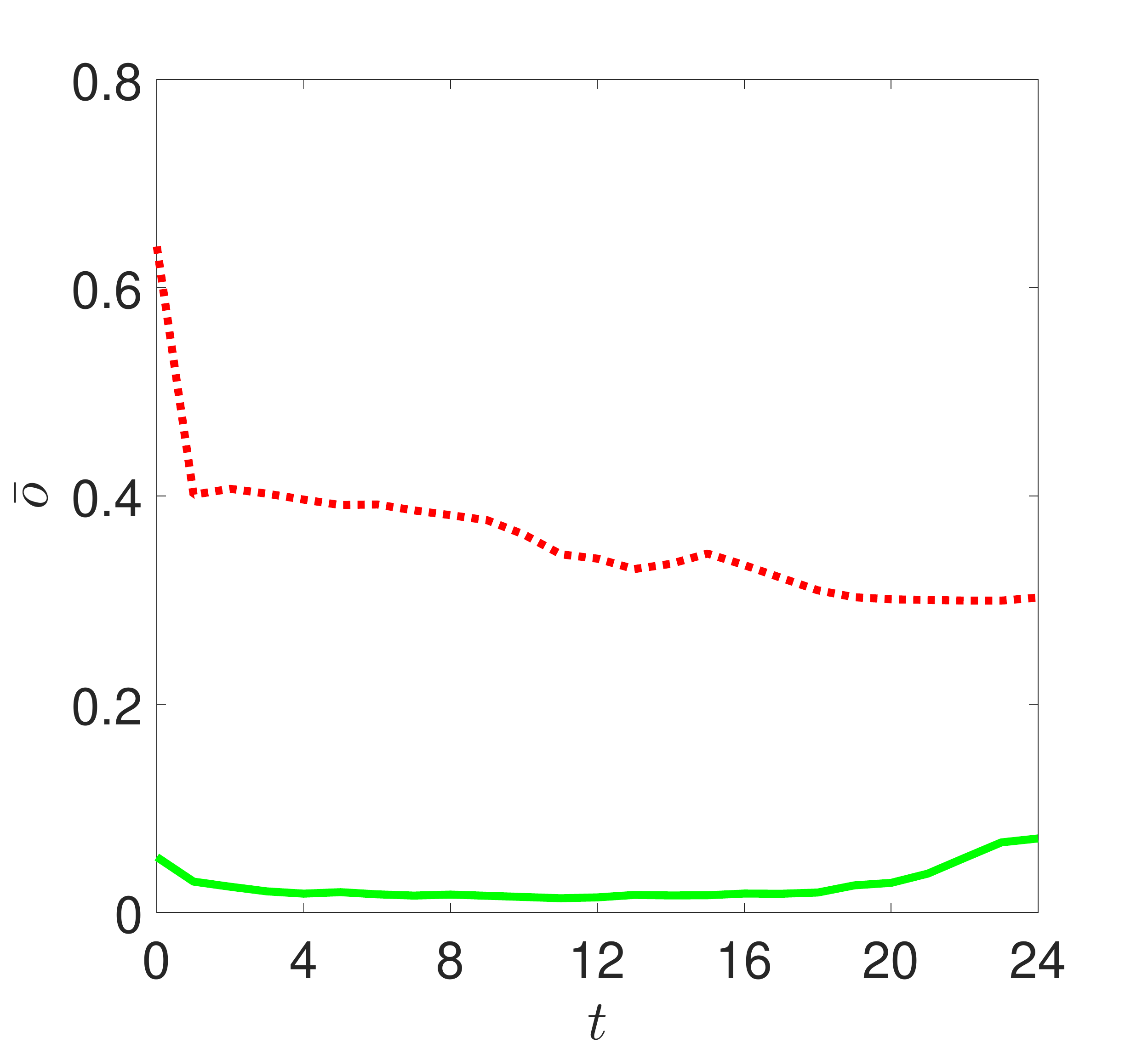,width = 0.64 \linewidth}}
\end{picture}
\end{center}
      \caption{Example 2: Time histories of average extent of detection when solely pursuing the nominal objective (in red dotted) and when pursuing both the nominal and the anti-detection objectives (in green solid).}
      \label{fig: EX2_O}
      \vspace{0.05in}
\end{figure}

It can be observed from Figs.~\ref{fig: EX2_path} and \ref{fig: EX2_O} that by executing the mission along a sub-optimal path in terms of the nominal reward, the ego agent can significantly reduce the extent of detection by the radar sensor.

\section{Conclusions}\label{sec:6}

In this paper, we defined the detection-averse optimal control problem for Markov decision processes. A value iteration approach and a receding-horizon optimization approach were proposed to solve the problem, where the latter had a better scalablity to larger-sized problems. Two examples were reported to illustrate the two approaches and the potential of the problem formulation for practical applications.

\bibliographystyle{ifacconf-harvard}
\bibliography{ref}

\section*{Appendix}

{\it Lemma 2:} For any $(x,o) \in X \times \Delta$ such that $o(x)>0$, $U_a(x,o) \neq \emptyset$.

{\it Proof:} Suppose $o(x)\!>\!0$. For any $y \in Y$, if $\sum_{x' \in X} q(y|x')$ $p(x'|x,\pi^*(x)) = \sum_{x' \in X} q(y|x')\, p_a(x'|x)> 0$,
then $\sum_{x'' \in X} q(y|x'') \big(\sum_{x' \in X} p_a (x''|x') o(x')\big) \ge \sum_{x'' \in X}$ $q(y|x'') p_a (x''|x) o(x) > 0$. Thus, $u = \pi^*(x) \in U_a(x,o)$. $\blacksquare$

Lemma~2 verifies that for an arbitrary model $(p,q,p_a)$, the set $U_a(x,o)$ of admissible actions may only be empty at the points $(x,o) \in X \times \Delta$ where $o(x) = 0$, which represent a null set of the space $X \times \Delta$.

\end{document}